%% file: p139a.tex
\documentclass[11pt,epsf]{article}
\usepackage{amsmath}
\usepackage{amsfonts}
\usepackage{amssymb}
\usepackage{graphicx}
\usepackage{color}

\newtheorem{theorem}{Theorem}

\newcommand {\dfn} {\stackrel{\Delta} {=}}

\newcommand {\dd} {\mbox{d}}

\newcommand {\bmu} {\mbox{\boldmath $\mu$}}

\newcommand {\br} {\mbox{\boldmath $r$}}

\newcommand {\bu} {\mbox{\boldmath $u$}}

\newcommand {\bE} {\mbox{\boldmath $E$}}

\newcommand{\calU}{{\cal U}}
\newcommand{\calV}{{\cal V}}
\newcommand{\calW}{{\cal W}}
\newcommand{\calX}{{\cal X}}
\newcommand{\calY}{{\cal Y}}

\begin{document}
\thispagestyle{empty}
\title{Data Processing Theorems and the Second Law of
Thermodynamics}
\author{Neri Merhav\thanks{This work was supported by the Israel Science
Foundation (ISF) grant no.\ 208/08.}}
\date{}
\maketitle

\begin{center}
Department of Electrical Engineering \\
Technion - Israel Institute of Technology \\
Haifa 32000, ISRAEL \\
\end{center}
\vspace{1.5\baselineskip}
\setlength{\baselineskip}{1.5\baselineskip}

\begin{abstract}
We draw relationships between the generalized data processing theorems of
Zakai and Ziv (1973 and 1975) and the dynamical version of the second law of thermodynamics,
a.k.a.\ the Boltzmann H--Theorem, which asserts that the Shannon entropy,
$H(X_t)$, pertaining to a finite--state Markov process $\{X_t\}$, is monotonically
non--decreasing as a function of time $t$, provided that the steady--state
distribution of this process is uniform across 
the state space (which is the case when the process designates an isolated
system). It turns out
that both the generalized data processing theorems and the Boltzmann
H--Theorem can be viewed as special cases of a more general principle
concerning the monotonicity (in time) of a certain generalized information
measure applied to a Markov process. This gives rise to a new look at the 
generalized data processing
theorem, which suggests to exploit certain degrees of freedom that
may lead to better bounds, for a given choice of
the convex function that defines the generalized mutual information.\\

\end{abstract}

\newpage
\section{Introduction}

In \cite{Csiszar72}, Csisz\'ar considered a generalized notion of the
divergence between two probability distributions,
a.k.a.\ the {\it f--divergence},
by replacing the negative logarithm function, of the classical divergence,
\begin{equation}
D(P_1\|P_2)=\int\dd x\cdot P_1(x)\left[-\log \frac{P_2(x)}{P_1(x)}\right],
\end{equation}
with a general convex function\footnote{Originally, this function was denoted
by $f$ in \cite{Csiszar72}, hence the name f--divergence.} $Q$, i.e.,
\begin{equation}
D_Q(P_1\|P_2)=\int\dd x\cdot P_1(x)\cdot Q\left(\frac{P_2(x)}{P_1(x)}\right).
\end{equation}
When the f--divergence was applied to the joint
distribution (in the role of $P_1$) and the product of marginals 
(in the role of $P_2$) of two random variables, it yielded
a generalized notion of mutual information, 
\begin{equation}
I^Q(X;Y)=\int\dd x\dd y\cdot P(x,y)\cdot Q\left(\frac{P(x)P(y)}{P(x,y)}\right)
=\int\dd x\dd y\cdot P(x,y)\cdot Q\left(\frac{P(y)}{P(y|x)}\right),
\end{equation}
which was shown in
\cite{Csiszar72} to obey a data
processing inequality, thus extending the well known data processing
inequality of the ordinary mutual 
information (see, e.g., \cite[Section 2.8]{CT06}).

The same ideas were introduced independently by Ziv and Zakai \cite{ZZ73},
with the primary motivation of using it to obtain sharper distortion bounds 
for classes of simple codes for 
joint source--channel coding (e.g., of block length $1$), as well as certain
situations of signal detection and estimation (see also \cite{Andelman74}).
The idea was to
define both a ``rate--distortion function,'' $R^Q(d)$ and a ``channel
capacity,'' $C^Q$,
by minimization and maximization (respectively) of the mutual information
pertaining to $Q$, and to derive a lower bound on the distortion $d$ from 
the data processing inequality 
\begin{equation}
R^Q(d)\le C^Q.
\end{equation}
In the sequel, this will be referred to as the 1973 version of the generalized
data processing theorem.
In a somewhat less well known work \cite{ZZ75}, Zakai and Ziv have
substantially further generalized their data processing theorems, so
as to apply an  even more general information measures, and
this will be referred to as the 1975 version.
This generalized information measure was in the form
\begin{align}
\label{zzim75}
I^Q(X;Y)&=\int\dd x\dd y\cdot P(x,y)\cdot
Q\left(\frac{\mu_1(x,y)}{P(x,y)},\ldots,\frac{\mu_k(x,y)}{P(x,y)}\right)\nonumber\\
&=\int\dd x\dd y\cdot P(x,y)\cdot
Q\left(\frac{\mu_1(y|x)}{P(y|x)},\ldots,\frac{\mu_k(y|x)}{P(y|x)}\right),
\end{align}
where $Q$ is now an arbitrary convex function of $k$ variables and
$\{\mu_i(x,y)\}$ are arbitrary positive measures (not necessarily probability
measures) that are defined consistently with the Markov conditions and
where $\mu_i(y|x)=\mu_i(x,y)/P(x)$. 
It was shown in \cite[Theorem 7.1]{ZZ75} that the
distortion bounds obtained from (\ref{zzim75}) are tight in the sense that
there always exist a convex function $Q$ and measures $\{\mu_i\}$ that would
yield the exact distortion pertaining to the optimum communication system, and
so, there is no room for improvement of this class of bounds.\footnote{This
result is non--constructive, however, in the sense that this choice of $Q$ and
$\{\mu_i\}$ depends on the optimum encoder and decoder.}

By setting $\mu_i(y|x)=P(y|x_i)$, $i=1,2,\ldots,k-1$, where $\{x_i\}$ are 
$k-1$ particular letters in the
alphabet of $X$, and $\mu_k(y|x)=P(y)$, 
they defined yet another generalized information
measure that satisfies the data processing theorem as
\begin{equation}
\bE\left\{Q\left(\frac{P(Y|X_1)}{P(Y|X)},\ldots,
\frac{P(Y|X_{k-1})}{P(Y|X)},\frac{P(Y)}{P(Y|X)}\right)\right\},
\end{equation}
where the expectation is taken w.r.t.\ the joint distribution
$$P(x_1,\ldots,x_{k-1},x,y)=P(x)P(y|x)P(x_1)P(x_2)\cdot\cdot\cdot P(x_{k-1}).$$
In both \cite{ZZ73} and \cite{ZZ75}, there are many examples how these data
processing inequalities can be used to improve on earlier distortion bounds.

The data processing theorems of Csisz\'ar and Zakai and Ziv form one aspect
of this work. The other aspect, which may seem unrelated at first glance
(but will nevertheless be shown here to be strongly related)
is the second law of thermodynamics, or more precisely, {\it Boltzmann's
H--theorem.} The second law of thermodynamics tells that in an isolated
physical system (i.e., when no energy flows in or out), the entropy cannot
decrease over time. Since one of the basic postulates of statistical physics
tells that all states of the system, which have the same energy, also have
the same probability in equilibrium, it follows that the
stationary 
(equilibrium) distribution of these states must be uniform, because all
accessible states must have the same energy when the system is isolated.
Indeed, if the state of this system is designated by a Markov
process, $\{X_t\}$ with a uniform stationary state distribution, the
Boltzmann H--theorem tells that the Shannon entropy of $X_t$, $H(X_t)$, cannot decrease
with $t$, which is a restatement of the second law.

We show, in this paper, that the generalized data processing theorems of
\cite{Csiszar72}, \cite{ZZ73}, and \cite{ZZ75} on the one hand, and the
Boltzmann H--theorem,
on the other hand,
are all special cases of a more general principle, which asserts that a
certain generalized information measure, applied to the underlying Markov
process must be a monotonic function of time.
This unified framework provides a new perspective
on the generalized data processing theorem.
Beyond the fact that this new perspective may be interesting on its own
right, it naturally suggests to exploit 
certain degrees of freedom of the Ziv--Zakai 
generalized mutual information that
may lead to better bounds, for a given choice of
the convex function that defines this generalized mutual information.
These additional degrees of freedom may be important, because the
variety of convex functions $\{Q\}$ which are convenient to work with, is
rather limited.
The fact that better bounds may indeed be obtained is demonstrated by an
example.

The outline of the remaining part of this paper is as follows.
In Section \ref{background}, we provide some background on Markov
processes with a slight physical flavor, which will include the
notion of detailed balance, global balance, as well as 
known results like the Boltzmann H--theorem, and its generalizations
to information measures other than the entropy.
In Section \ref{main}, we relate 
the generalized version of the Boltzmann H--theorem
and the generalized data processing theorems and formalize the uniform
framework that supports both. This is done, first for the 1973 version
\cite{ZZ73} of
the Ziv--Zakai data processing theorem (along with an example), and
then for the 1975 version by Zakai and Ziv \cite{ZZ75}.
Finally, in Section \ref{summary}, we summarize and conclude.

\section{Background}
\label{background}

\subsection{Detailed Balance and Global Balance}

Many dynamical models of a physical system describe the microscopic
state (or microstate, for short) of this system
as a Markov process, $\{X_t\}$, either in discrete time
or in continuous time.
In this section, we discuss a few properties of these processes
as well as the evolution of information measures associated with them, like
entropy, divergence and more.

We begin with an isolated system in continuous time, which
is not necessarily assumed to have reached yet 
its stationary distribution pertaining to equilibrium. Let us suppose
that the state $X_t$
may take on values in a finite set $\calX$. For
$x,x'\in\calX$, let us define the state transition rates
\begin{equation}
W_{xx'}=\lim_{\delta\to
0}\frac{\mbox{Pr}\{X_{t+\delta}=x'|X_t=x\}}{\delta}~~~~x'\ne x
\end{equation}
which means, in other words,
\begin{equation}
\mbox{Pr}\{X_{t+\delta}=x'|X_t=x\}=W_{xx'}\cdot \delta
+o(\delta). 
\end{equation}
Denoting
\begin{equation}
P_t(x)=\mbox{Pr}\{X_t=x\}, 
\end{equation}
it is easy to see that
\begin{equation}
P_{t+dt}(x)=\sum_{x'\ne x}P_t(x')W_{x'x}\mbox{d}t+P_t(x)\left(1-\sum_{x'\ne
x}W_{xx'}\mbox{d}t\right),
\end{equation}
where the first sum describes the probabilities of all possible transitions
from other states to
state $x$ and the second term describes the probability of not leaving state
$x$. Subtracting $P_t(x)$ from both sides and dividing by $\mbox{d}t$, we
immediately obtain the following set of differential equations:
\begin{equation}
\frac{\mbox{d}P_t(x)}{\mbox{d}t}=
\sum_{x'}[P_t(x')W_{x'x}-P_t(x)W_{xx'}],~~~x\in\calX,
\end{equation}
where $W_{xx}$ is defined in an arbitrary manner, e.g., $W_{xx}\equiv 0$ for
all $x\in\calX$. In the physics terminology (see, e.g.,
\cite{Kittel58},\cite{Reif65}),
these equations are called the {\it master equations}.\footnote{Note that the
master equations apply in discrete time too, provided that the derivative at
the l.h.s.\ is replaced by a simple difference, $P_{t+1}(x)-P_t(x)$, and
$\{W_{xx'}\}$ are replaced one--step state transition probabilities.}
When the process
reaches stationarity, i.e., for all $x\in\calX$, $P_t(x)$ converge to some
$P(x)$ that is
time--invariant, then
\begin{equation}
\sum_{x'}[P(x')W_{x'x}-P(x)W_{xx'}]=0,~~~\forall~x\in\calX .
\end{equation}
This situation is called
{\it global balance} or {\it steady state}. When the physical system
under discussion is isolated, namely, no energy flows into the system or
out, the steady state distribution must be uniform across all states, because
all accessible states must be of the same energy and the equilibrium
probability of each state depends solely on its energy. Thus,
in the case of an isolated system,
$P(x)=1/|\calX|$ for all $x\in\calX$.
From quantum mechanical considerations, as well as considerations
pertaining to time reversibility in the microscopic level,\footnote{Consider,
for example, an isolated system
of moving particles of mass $m$ and position vectors
$\{\br_i(t)\}$, obeying the differential
equations $m\mbox{d}^2\br_i(t)/\mbox{d}t^2=\sum_{j\ne i}F(\br_j(t)-\br_i(t))$,
$i=1,2,\ldots,n$, ($F(\br_j(t)-\br_i(t))$ being mutual interaction
forces),
which remain valid if the time variable $t$ is replaced by $-t$ since
$\mbox{d}^2\br_i(t)/\mbox{d}t^2=\mbox{d}^2\br_i(-t)/\mbox{d}(-t)^2$.}
it is customary to
assume $W_{xx'}=W_{x'x}$ for all pairs $\{x,x'\}$. We then observe that, not
only do $\sum_{x'}[P(x')W_{x'x}-P(x)W_{xx'}]$ all vanish, but moreover, each
individual term in
this sum vanishes, as
\begin{equation}
P(x')W_{x'x}-P(x)W_{xx'}=\frac{1}{|\calX|}(W_{x'x}-W_{xx'})=0.
\end{equation}
This property is called {\it detailed balance}, which is stronger than global
balance, and it means equilibrium, which is stronger than steady state. While
both steady--state and equilibrium refer to situations of time--invariant
state probabilities $\{P(x)\}$, a steady--state still allows cyclic ``flows of
probability.'' For example, a Markov process with cyclic deterministic
transitions $1\to 2\to 3\to 1\to 2\to 3\to \cdot\cdot\cdot$ is in steady state
provided that the probability distribution of the initial state is uniform
$(1/3,1/3,1/3)$, however, the cyclic flow among the states is in one
direction. On the other hand, in detailed balance ($W_{xx'}=W_{x'x}$ for an
isolated system),
which is equilibrium, there
is no net flow in any cycle of states. All the net cyclic
probability fluxes vanish, and
therefore, time reversal would not change the probability law, that is,
$\{X_{-t}\}$ has the same probability law as $\{X_t\}$ (see \cite[Sect.\
1.2]{Kelly79}). For example,
if $\{Y_t\}$ is a Bernoulli process, taking values equiprobably in
$\{-1,+1\}$, then $X_t$ defined
recursively by
\begin{equation}
X_{t+1}=(X_t+Y_t)\mbox{mod} K,
\end{equation}
has a symmetric state--transition probability matrix $W$, a uniform stationary
state distribution, and it satisfies detailed balance.

\subsection{Monotonicity of Information Measures}
\label{monotonicity}

Returning to the case where the process
$\{X_t\}$ pertaining to our isolated system has not necessarily reached
equilibrium,
let us take a look at the entropy of the state
\begin{equation}
H(X_t)=-\sum_{x\in\calX}P_t(x)\log
P_t(x).
\end{equation}
The Boltzmann H--theorem (see, e.g., \cite[Chap.\ 7]{Beck76}, \cite[Sect.\
3.5]{Kardar07}, \cite[pp.\ 171--173]{Kittel58}
\cite[pp.\ 624--626]{Reif65}) asserts
that $H(X_t)$ is monotonically non--decreasing. This result is a
restatement of the second law of thermodynamics,
which tells that the entropy of an isolated system
cannot decrease with time. To see why this is true,
we next show that detailed balance implies
\begin{equation}
\frac{\mbox{d}H(X_t)}{\mbox{d}t}\ge 0,
\end{equation}
where for convenience, we denote
$\mbox{d}P_t(x)/\mbox{d}t$ by $\dot{P}_t(x)$. Now,
\begin{align}
\frac{\mbox{d}H(X_t)}{\mbox{d}t}&=-\sum_x[\dot{P}_t(x)\log
P_t(x)+\dot{P}_t(x)]\nonumber\\
&=-\sum_x\dot{P}_t(x)\log
P_t(x)\nonumber\\
&=-\sum_x\sum_{x'}W_{x'x}[P_t(x')-P_t(x)]\log
P_t(x))\nonumber\\
&=-\frac{1}{2}\sum_{x,x'}W_{x'x}[P_t(x')-P_t(x)]\log P_t(x)-\nonumber\\
& \frac{1}{2}\sum_{x,x'}W_{x'x}[P_t(x)-P_t(x')]\log
P_t(x')\nonumber\\
&=\frac{1}{2}\sum_{x,x'}W_{x'x}[P_t(x')-P_t(x)]\cdot[\log P_t(x')-\log
P_t(x)]\nonumber\\
&\ge 0,
\end{align}
where in the second line we used the fact that $\sum_x\dot{P}_t(x)=0$,
in the third line we used detailed balance ($W_{xx'}=W_{x'x}$),
and the last inequality is due to the increasing monotonicity of the
logarithmic
function: the product 
$[P_t(x')-P_t(x)]\cdot[\log P_t(x')-\log P_t(x)]$ cannot be
negative for any pair $(x,x')$, as the two factors of this product are either
both negative,
both zero, or
both positive. Thus, $H(X_t)$ cannot decrease with time.

The H--theorem has a discrete--time analogue: If a finite--state Markov
process has a symmetric transition probability matrix (which is the
discrete--time counterpart of the above detailed balance property), which
means that the stationary
state distribution is uniform, then $H(X_t)$ is a monotonically
non--decreasing sequence.

A well--known paradox, in this context, is associated with the notion of the
{\it arrow of time.} On the one hand, we are talking about time--reversible
processes, obeying detailed balance, but on the other hand, the increase of
entropy suggests that there is asymmetry between the two possible directions
that the time axis can be exhausted, the forward direction and the backward
direction. If we go back in time, the entropy would decrease. So is there an
arrow of time? This paradox was resolved, by Boltzmann himself, once he
made the clear
distinction between equilibrium and non--equilibrium situations: The notion
of time reversibility is associated with equilibrium, where the process
$\{X_t\}$ is stationary. On the other hand, the increase of entropy is a
result that belongs to the non--stationary regime, where the process is on its
way to stationarity and equilibrium.
In the latter case, the system has been
initially prepared in a non--equilibrium situation.
Of course, when the process is stationary, $H(X_t)$ is fixed and there
is no contradiction.

\begin{figure}[ht]
\hspace*{2cm}\input{mm1.pstex_t}
\caption{\small State transition diagram of an M/M/1 queue.}
\label{mm1}
\end{figure}
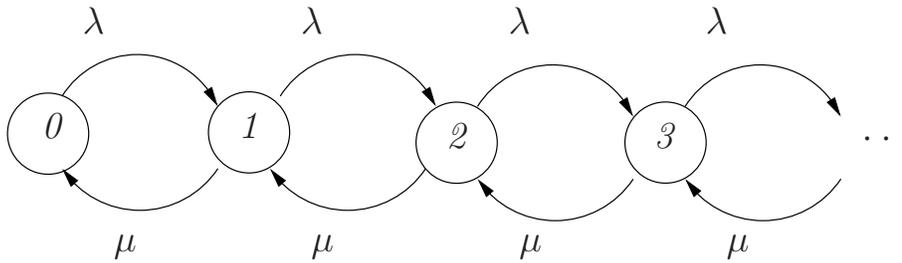

So far we discussed the property of detailed balance only for an isolated
system, where the stationary state distribution is the uniform distribution.
How is the property of detailed balance 
defined when the stationary distribution is
non--uniform?
For a general Markov process, whose steady state--distribution is not
necessarily uniform, the condition of detailed balance, which means
time--reversibility \cite{Kelly79}, reads
\begin{equation}
P(x)W_{xx'}=P(x')W_{x'x},
\end{equation}
in the continuous--time case. In the discrete--time case (where $t$ takes on positive
integer values only), 
it is defined by a similar equation, except that $W_{xx'}$ and
$W_{x'x}$ are replaced by the corresponding
one--step state transition probabilities, i.e.,
\begin{equation}
P(x)P(x'|x)=P(x')P(x|x'),
\end{equation}
where 
\begin{equation}
P(x'|x)\dfn \mbox{Pr}\{X_{t+1}=x'|X_t=x\}.
\end{equation}
The physical interpretation is that now our system is (a small)
part of a much larger isolated system, which obeys detailed balance w.r.t.\ the
uniform equilibrium distribution, as before.
A well known example of a process that obeys detailed balance in its more
general form
is the M/M/1 queue with an arrival rate $\lambda$ and service rate $\mu$
($\lambda < \mu$). Here, since all states are arranged along a line, with
bidirectional transitions between neighboring states only (see Fig.\
\ref{mm1}), there cannot be
any cyclic probability flux. The steady--state distribution is well--known
to be geometric
\begin{equation}
P(x)=\left(1-\frac{\lambda}{\mu}\right)
\cdot\left(\frac{\lambda}{\mu}\right)^x, ~~~~~~x=0,1,2,\ldots,
\end{equation}
which indeed satisfies the detailed
balance $P(x)\lambda=P(x+1)\mu$ for all $x$.
Thus, the Markov process $\{X_t\}$,
designating the number of customers in the queue at time $t$,
is time--reversible.

For the sake of simplicity, from this point onward, our discussion will focus
almost exclusively on discrete--time Markov processes, but the results to
be stated, will hold for continuous--time Markov processes as well.
We will continue to denote by $P_t(x)$ the probability of $X_t=x$, except that
now $t$ will be limited to take on integer values only. The one--step state
transition probabilities will be denoted by $\{P(x'|x)\}$, as mentioned
earlier.

How does the H--theorem extend to situations where the stationary state
distribution is not uniform? In \cite[p.\ 82]{CT06}, it is shown (among other
things) that
the divergence,
\begin{equation}
\label{divdecrease}
D(P_t\|P)=\sum_{x\in\calX}P_t(x)\log\frac{P_t(x)}{P(x)},
\end{equation}
where $P=\{P(x),~x\in\calX\}$ is a stationary state 
distribution, is a monotonically non--increasing function of
$t$. Does this result have a physical interpretation, like the H--theorem and
the second law of thermodynamics? When it comes to non--isolated systems,
where the steady state distribution is non--uniform, the extension of the
second law of thermodynamics, replaces the principle of increase of entropy by
the principle of decrease of free energy, or equivalently, the decrease of
the difference between the free energy at time $t$ and the free energy in
equilibrium. The information--theoretic counterpart of this free energy
difference is the divergence $D(P_t\|P)$
(see, e.g., \cite{Bagci07}). Thus, the monotonic decrease of $D(P_t\|P)$
has a simple physical interpretation of free energy decrease, which 
is the natural extension of the entropy increase. 
Indeed, particularizing this to the case where
$P$ is the uniform distribution (as in an isolated system), then
\begin{equation}
D(P_t\|P)=\log|\calX|-H(X_t),
\end{equation}
which means that the decrease of the divergence is equivalent to the increase of
entropy, as before. However, here the result is more general than the
H--theorem from an additional aspect: It does not require detailed balance.
It only requires the existence of the stationary state distribution.
Note that even in the earlier case, of an isolated system,
detailed balance, which means symmetry of the state transition probability
matrix ($P(x'|x)=P(x|x')$), is a stronger requirement than uniformity of the
stationary state distribution, as the latter requires merely that
the matrix $\{P(x'|x)\}$ would be doubly stochastic, i.e.,
$\sum_xP(x|x')=\sum_xP(x'|x)=1$ for all $x'\in\calX$, which is weaker than
symmetry of the matrix itself.
The results shown in \cite{CT06} are, in fact, somewhat more general:
Let $P_t=\{P_t(x)\}$ and $P_t'=\{P_t'(x)\}$ be two time--varying
state--distributions
pertaining to the same Markov chain, but induced by two different initial
state distributions, $\{P_0(x)\}$ and $\{P_0'(x)\}$, respectively. Then $D(P_t\|P_t')$ is
monotonically non--increasing. This is easily seen as follows:
\begin{align}
\label{covermono}
D(P_t\|P_t')&=\sum_xP_t(x)\log\frac{P_t(x)}{P_t'(x)}\nonumber\\
&=\sum_{x,x'}P_t(x)P(x'|x)
\log\frac{P_t(x)P(x'|x)}{P_t'(x)P(x'|x)}\nonumber\\
&=\sum_{x,x'}P(X_t=x,~X_{t+1}=x')
\log\frac{P(X_t=x,~X_{t+1}=x')}{P'(X_t=x,~X_{t+1}=x')}\nonumber\\
&\ge D(P_{t+1}\|P_{t+1}')
\end{align}
where the last inequality follows from the data processing theorem of the
divergence: the divergence between two joint distributions of
$(X_t,X_{t+1})$
is never smaller than the
divergence between corresponding marginal distributions of $X_{t+1}$.
Another interesting special case of this result is obtained if we now take 
the first argument of the divergence to the a stationary state distribution:
This will mean that $D(P\|P_t)$ is also monotonically non--increasing.

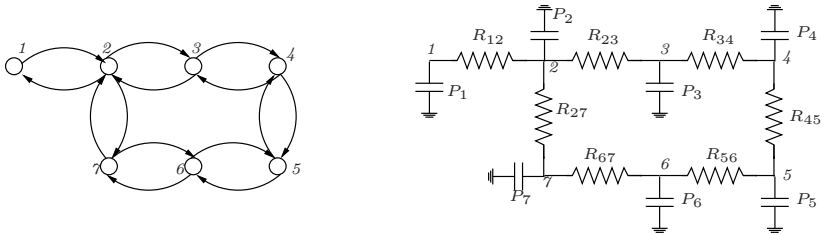
\begin{figure}[ht] 
\hspace*{2cm}\input{circuit.pstex_t}
\caption{\small State transition diagram of a Markov chain
(left part) and the electric circuit that emulates the
dynamics of $\{P_t(x)
\}$ (right part).}
\label{circuit}
\end{figure}

In \cite[Theorem 1.6]{Kelly79}, there is a further extension of
all the above monotonicity results, where the
ordinary divergence is actually replaced by the f--divergence (though
the relation to the f--divergence is not mentioned in \cite{Kelly79}):
If $\{X_t\}$
is a Markov process with a given state transition probability matrix
$\{P(x'|x)\}$, then the function
\begin{equation}
U(t)=D_Q(P\|P_t)=\sum_{x\in\calX}P(x)\cdot Q\left(\frac{P_t(x)}{P(x)}\right)
\end{equation}
is monotonically non--increasing, provided that
$Q$ is convex. Moreover, $U(t)$ monotonically strictly decreasing if
$Q$ is strictly convex and $\{P_t(x)\}$ is not identical to $\{P(x)\}$).
To see why this is true,
define the backward
transition probability matrix by
\begin{equation}
\tilde{P}(x|x')=\frac{P(x)P(x'|x)}{P(x')}. 
\end{equation}
Obviously,
\begin{equation}
\sum_x\tilde{P}(x|x')=1 
\end{equation}
for all
$x'\in\calX$, and so,
\begin{equation}
\frac{P_{t+1}(x)}{P(x)}=\sum_{x'}\frac{P_t(x')P(x|x')}{P(x)}=
\sum_{x'}\frac{\tilde{P}(x'|x)P_t(x')}{P(x')}.
\end{equation}
By the convexity of $Q$:
\begin{align}
U(t+1)&=\sum_xP(x)\cdot Q\left(\frac{P_{t+1}(x)}{P(x)}\right)\nonumber\\
&=\sum_xP(x)\cdot
Q\left(\sum_{x'}\tilde{P}(x'|x)\frac{P_t(x')}{P(x')}\right)\nonumber\\
&\le\sum_x\sum_{x'}P(x)\tilde{P}(x'|x)\cdot
Q\left(\frac{P_t(x')}{P(x')}\right)\nonumber\\
&=\sum_x\sum_{x'}P(x')P(x|x')\cdot
Q\left(\frac{P_t(x')}{P(x')}\right)\nonumber\\
&=\sum_{x'}P(x')\cdot Q\left(\frac{P_t(x')}{P(x')}\right)=U(t).
\end{align}
Now, a few interesting choices of the function $Q$ may be considered:
As proposed in \cite[p.\ 19]{Kelly79}, for $Q(u)=u\ln u$, we have
$U(t)=D(P_t\|P)$, and we are back to the aforementioned result in \cite{CT06}.
Another interesting choice is $Q(u)=-\ln u$,
which gives $U(t)=D(P\|P_t)$. Thus, the monotonicity of $D(P\|P_t)$ is also
obtained as a special case.\footnote{We are not yet in a position to obtain the
monotonicity of $D(P_t\|P_t')$ as a special case of the monotonicity of
$D_Q(P\|P_t)$. This will require a slight further extension of this
information measure, to be carried out later on.}
Yet another choice is $Q(u)=-u^s$, where $s\in[0,1]$ is a
parameter. This would yield the increasing monotonicity of
$\sum_x P^{1-s}(x)P_t^s(x)$, a `metric' that plays a role in
the theory of asymptotic
exponents of error probabilities pertaining to the optimum
likelihood ratio test between two probability distributions \cite[Chapter
3]{VO79}. In particular,
the choice $s=1/2$ yields balance between the two kinds of error and it is
intimately related to the Bhattacharyya distance.
In the case of detailed balance, there is another physical 
interpretation of the approach to equilibrium and the growth of $U(t)$
\cite[p.\ 20]{Kelly79}: Returning, for a moment, to the realm of
continuous--time Markov processes,
we can write the master equations as follows:
\begin{equation}
\frac{\mbox{d}P_t(x)}{\mbox{d}t}=\sum_{x'}\frac{1}{R_{xx'}}
\left[\frac{P_t(x')}{P(x')}-\frac{P_t(x)}{P(x)}\right]
\end{equation}
where $R_{xx'}=[P(x')W_{x'x}]^{-1}=[P(x)W_{xx'}]^{-1}$. Imagine now an
electrical
circuit where the indices $\{x\}$ designate the various nodes. Nodes $x$ and
$x'$ are connected by
a wire with resistance $R_{xx'}$ and every node $x$ is grounded via a
capacitor
with capacitance $P(x)$ (see Fig.\ \ref{circuit}).
If $P_t(x)$ is the charge at node $x$ at time $t$,
then the master equations are the Kirchoff equations of the currents at each
node in the
circuit. Thus, the way in which probability spreads across the states is
analogous to the way charge spreads across the circuit and probability fluxes
are now analogous to electrical currents.
If we now choose
$Q(u)=\frac{1}{2}u^2$, then
\begin{equation} 
U(t)=\frac{1}{2}\sum_x\frac{P_t^2(x)}{P(x)}, 
\end{equation} 
which means that the energy stored in the capacitors 
dissipates as heat in the wires until the system 
reaches equilibrium, where all nodes have the same 
potential, $P_t(x)/P(x)=1$, and hence detailed 
balance corresponds to the situation where 
all individual currents vanish (not only 
their algebraic sum).  

We have seen, in the above examples, that various choices of the function $Q$
yield various f--divergences, or `metrics',
between $\{P(x))\}$ and $\{P_t(x)\}$,
which are both
marginal distributions of a single symbol $x$. 
What about joint distributions
of two or more symbols? Consider, for example, the function
\begin{equation}
J(t)=\sum_{x,x'}P(X_0=x,X_t=x')\cdot
Q\left(\frac{P(X_0=x)P(X_t=x')}{P(X_0=x,X_t=x')}\right),
\end{equation}
where $Q$ is convex as before. Here, by the same token, $J(t)$ is the
f--divergence
between the joint probability distribution $\{P(X_0=x,X_t=x')\}$ and the
product of marginals $\{P(X_0=x)P(X_t=x')\}$, namely, it 
is the generalized mutual information of \cite{Csiszar72},\cite{ZZ73}, and
\cite{ZZ75}, as mentioned in the Introduction.
Now, using
a similar chain of inequalities as before, we get the non--decreasing
monotonicity of $J(t)$ as follows:
\begin{align}
J(t)&=\sum_{x,x',x''}P(X_0=x,X_t=x',X_{t+1}=x'')\times\nonumber\\
&  Q\left(\frac{P(X_0=x)P(X_t=x')}{P(X_0=x,X_t=x')}
\cdot\frac{P(X_{t+1}=x''|X_t=x')}{P(X_{t+1}=x''|X_t=x')}\right)\nonumber\\
&=\sum_{x,x''}P(X_0=x,X_{t+1}=x'')\sum_{x'}P(X_t=x'|X_0=x,X_{t+1}=x'')\times\nonumber\\
& Q\left(\frac{P(X_0=x)P(X_t=x',X_{t+1}=x'')}
{P(X_0=x,X_t=x',X_{t+1}=x'')}\right)\nonumber\\
&\le\sum_{x,x''}P(X_0=x,X_{t+1}=x'')\cdot
Q\left(\sum_{x'} P(X_t=x'|X_0=x,X_{t+1}=x'')\times\right.\nonumber\\
&  \left.\frac{P(X_0=x)P(X_t=x',X_{t+1}=x'')}
{P(X_0=x,X_t=x',X_{t+1}=x'')}\right)\nonumber\\
&=\sum_{x,x''}P(X_0=x,X_{t+1}=x'')
Q\left(\sum_{x'}
\frac{P(X_0=x)P(X_t=x',X_{t+1}=x'')}
{P(X_0=x,X_{t+1}=x'')}\right)\nonumber\\
&=\sum_{x,x''}P(X_0=x,X_{t+1}=x'')\cdot Q\left(
\frac{P(X_0=x)P(X_{t+1}=x'')}
{P(X_0=x,X_{t+1}=x'')}\right)\nonumber\\
&=J(t+1).
\end{align}
This time, we assumed only the Markov property of $(X_0,X_t,X_{t+1})$ (not even homogeneity). 
This is, in fact, nothing but 
the 1973 version of the generalized
data processing theorem of Ziv and
Zakai \cite{ZZ73}, which was mentioned in the Introduction.

\section{A Unified Framework}
\label{main}

In spite of the general resemblance (via the notion of the f--divergence),
the last monotonicity result, concerning $J(t)$, and the monotonicity of $D(P_t\|P_t')$,
do not seem, at first glance, to fall in the framework 
of the monotonicity of the f--divergence $D_Q(P\|P_t)$.
This is because in the latter, there is an additional dependence on a stationary state
distribution that appears neither in $D(P_t\|P_t')$ nor in $J(t)$. However,
two simple observations
can put them both in the framework of the monotonicity of
$D_Q(P\|P_t)$. 

The first observation
is that the monotonicity of $U(t)=D_Q(P\|P_t)$ continues to hold (with a
straightforward extension of the proof) if $P_t(x)$ is extended to be a vector
of time varying state distributions
$(P_t^{1}(x),P_t^{2}(x),\ldots,P_t^{k}(x))$, and $Q$
is taken to be a convex function of
$k$ variables.
Moreover, each component $P_t^{i}(x)$ does not have to be
necessarily a probability distribution. It can be any function
$\mu_t^{i}(x)$ that satisfies
the recursion
\begin{equation}
\mu_{t+1}^{i}(x)=\sum_{x'} \mu_t^{i}(x')P(x|x'),~~~~~1\le i\le k.
\end{equation}
Let us then denote
$\bmu_t(x)=(\mu_t^{1}(x),\mu_t^{2}(x),\ldots,\mu_t^{k}(x))$
and assume that $Q$ is jointly convex in all its $k$ arguments.
Then the redefined function
\begin{align}
U(t)&=\sum_{x\in\calX} P(x)\cdot
Q\left(\frac{\bmu_t(x)}{P(x)}\right)\nonumber\\
&=\sum_{x\in\calX} P(x)\cdot
Q\left(\frac{\mu_t^{1}(x)}{P(x)},\ldots,\frac{\mu_t^{k}(x)}{P(x)}\right)
\end{align}
is monotonically non--increasing with $t$.

The second observation is rooted in convex analysis,
and it is related to the notion of the perspective of a convex
function and its convexity property \cite{BV04}. 
Here, a few words of background are in order. Let
$Q(\bu)$ be a convex function of the vector $\bu=(u_1,\ldots,u_k)$
and let $v > 0$ be an additional variable. Then, the function
\begin{equation}
\tilde{Q}(v,u_1,u_2,\ldots,u_k)\dfn v\cdot
Q\left(\frac{u_1}{v},\frac{u_2}{v},\ldots,\frac{u_k}{v}\right)
\end{equation}
is called the {\it perspective function} of $Q$. A well--known property of
the perspective operation is conservation of convexity, in other words, if $Q$
is convex in $\bu$, then $\tilde{Q}$ is convex in
$(v,\bu)$. The proof of this fact,
which is straightforward, can be found, for example, in
\cite[p.\ 89, Subsection 3.2.6]{BV04} (see also \cite{DM07}) and it is brought
here for the sake of completeness: Letting $\lambda_1$ and $\lambda_2$ be two
non--negative numbers summing to unity and letting $(v_1,\bu_1)$ and
$(v_2,\bu_2)$ be given, then
\begin{align}
\tilde{Q}(\lambda_1(v_1,\bu_1)+\lambda_2(v_2,\bu_2))
&= (\lambda_1v_1+\lambda_2v_2)\cdot 
Q\left(\frac{\lambda_1\bu_1+\lambda_2\bu_2}{\lambda_1v_1+\lambda_2v_2}\right)
\nonumber\\
&= (\lambda_1v_1+\lambda_2v_2)\cdot Q\left(
\frac{\lambda_1v_1}{\lambda_1v_1+\lambda_2v_2}\cdot\frac{\bu_1}{v_1}+ 
\frac{\lambda_2v_2}{\lambda_1v_1+\lambda_2v_2}\cdot\frac{\bu_2}{v_2}\right)\nonumber\\
&\le \lambda_1v_1Q\left(\frac{\bu_1}{v_1}\right)+
\lambda_2v_2Q\left(\frac{\bu_2}{v_2}\right)\nonumber\\
&=\lambda_1\tilde{Q}(v_1,\bu_1)+\lambda_2\tilde{Q}(v_2,\bu_2).
\end{align}
Putting these two observations together, we can now state the following
result:

\begin{theorem}
\label{hdpt}
Let
\begin{equation}
V(t)=\sum_x
\mu_t^{0}(x)Q\left(\frac{\mu_t^{1}(x)}{\mu_t^{0}(x)},
\frac{\mu_t^{2}(x)}{\mu_t^{0}(x)},
\ldots,\frac{\mu_t^{k}(x)}{\mu_t^{0}(x)}\right),
\end{equation}
where $Q$ is a convex function of
$k$ variables and
$\{\mu_t^{i}(x)\}_{i=0}^k$ are arbitrary functions
that satisfy the recursion
\begin{equation}
\mu_{t+1}^{i}(x)=\sum_{x'}\mu_t^{i}(x')P(x|x'),~~~~~~i=0,1,2,\ldots,k,
\end{equation}
and where $\mu_t^{0}(x)$ is moreover strictly positive.
Then, $V(t)$ is a monotonically non--increasing function of $t$.
\end{theorem}

Using the above mentioned observations, the proof of Theorem \ref{hdpt}
is straightforward: Letting $P$ be a stationary state distribution of
$\{X_t\}$, we have:
\begin{align}
V(t)&=\sum_x
\mu_t^{0}(x)Q\left(\frac{\mu_t^{1}(x)}{\mu_t^{0}(x)},
\frac{\mu_t^{2}(x)}{\mu_t^{0}(x)},
\ldots,\frac{\mu_t^{k}(x)}{\mu_t^{0}(x)}\right)\nonumber\\
&=\sum_x
P(x)\cdot\frac{\mu_t^{0}(x)}{P(x)}Q\left(\frac{\mu_t^{1}(x)/P(x)}{\mu_t^{0}(x)/P(x)},\ldots,
\frac{\mu_t^{k}(x)/P(x)}{\mu_t^{0}(x)/P(x)}\right)\nonumber\\
&=\sum_xP(x)\tilde{Q}\left(\frac{\mu_t^{0}(x)}{P(x)},\frac{\mu_t^{1}(x)}{P(x)},\ldots,
\frac{\mu_t^{k}(x)}{P(x)}\right).
\end{align}
Since $\tilde{Q}$ is the perspective of the convex function $Q$,
then it is convex as well, and so, the monotonicity of $V(t)$ follows from the
first observation above. 
It is now readily seen that both $D(P_t\|P_t')$ and $J(t)$ are special cases
of $V(t)$ and hence we have covered all special cases seen thus far under the
umbrella of the more general information functional $V(t)$. 

It is important to observe that the same idea exactly can be applied,
first of all, to
the 1973 version of the Ziv--Zakai data processing theorem (regardless of the
above described monotonicity results concerning Markov processes): Consider the generalized
mutual information functional
\begin{equation}
J^Q(X;Y)\dfn\sum_{x,y}\mu_0(x,y)Q\left(\frac{\mu_1(x,y)}{\mu_0(x,y)}\right),
\end{equation}
where $\mu_0(x,y)> 0$ and $\mu_1(x,y)$ are arbitrary functions that are
consistent with the Markov conditions, i.e., for any Markov chain $X\to Y\to
Z$, these functions satisfy 
\begin{equation}
\mu_i(x,z)=\sum_y\mu_i(x,y)P(z|y)=\sum_y\mu_i(y,z)P(x|y),~~~~ i=0,1.
\end{equation}
Then, $J^Q(X;Y)$ satisfies a data processing inequality, because, again
\begin{align}
J^Q(X;Y)&=\sum_{x,y}P(x,y)\cdot\frac{\mu_0(x,y)}{P(x,y)}
Q\left(\frac{\mu_1(x,y)/P(x,y)}{\mu_0(x,y)/P(x,y)}\right)\nonumber\\
&=\sum_{x,y}P(x,y)\tilde{Q}\left(
\frac{\mu_0(x,y)}{P(x,y)},\frac{\mu_1(x,y)}{P(x,y)}\right),
\end{align}
which is a Zakai--Ziv information functional of the 1975 version \cite{ZZ75}
and hence it satisfies a data processing inequality.

What functions, $\mu_0(x,y)$ and $\mu_1(x,y)$, can be consistent with the
Markov conditions? Two such functions are, of course, $\mu_0(x,y)=P(x,y)$ and
$\mu_1(x,y)=P(x)P(y)$, 
which bring us back to the 1973 Ziv--Zakai information measure.
We can, of course, swap their roles and obtain a generalized
version of the lautum information \cite{PV08}, which is also known to
satisfy a data processing inequality. For additional options,
let us consider a communication system, operating
on single symbols (block length 1), 
where the source symbol $u$ is mapped into a channel input $x=f(u)$,
by a deterministic encoder $f$, which is then fed into the channel
$P(y|x)$, and the channel output $y$
is in turn mapped into the reconstruction symbol $v=g(y)$.
As is argued in \cite{ZZ75}, the function $\mu(u,y)=P(u)P(y|u_0)$ is
consistent with the Markov conditions for any given source symbol $u_0$.
Indeed, since the encoder is assumed 
deterministic, $P(y|u_0)=P(y|f(u_0))=P(y|x_0)$, and
it is easily seen that
\begin{equation}
\mu(u,v)=P(u)P(v|u_0)=\sum_y P(u)P(y|u_0)P(v|y)=\sum_y\mu(u,y)P(v|y)
\end{equation}
and
\begin{align}
\mu(u,y)&=P(u)P(y|u_0)\nonumber\\
&=\sum_x P(u|x)P(x)P(y|u_0)\nonumber\\
&=\sum_xP(u|x)P(x)P(y|x_0)=
\sum_xP(u|x)\mu(x,y).
\end{align}
Of course, every linear combination of all these functions is also 
consistent with the Markov conditions. Thus, we can take
\begin{equation}
\mu_0(x,y)=s_0P(x,y)+\sum_{x_i\in\calX}s_i P(x)P(y|x_i)
\end{equation}
and
\begin{equation}
\mu_1(x,y)=t_0P(x,y)+\sum_{x_i\in\calX}t_i P(x)P(y|x_i),
\end{equation}
where $\{s_i\}$ and $\{t_i\}$ are the (arbitrary) coefficients of these linear
combinations (with the limitation that $s_i \ge 0$ for all $i$, with at least one $s_i > 0$).
Thus, we may define
\begin{equation}
J^Q(X;Y)=\sum_{x,y}\left[
s_0P(x,y)+\sum_{x_i\in\calX}s_i P(x)P(y|x_i)\right]\cdot Q\left(
\frac{t_0P(x,y)+\sum_{x_i\in\calX}t_i P(x)P(y|x_i)}
{s_0P(x,y)+\sum_{x_i\in\calX}s_i P(x)P(y|x_i)}\right),
\end{equation}
or, equivalently,
\begin{equation}
\label{withoutexp}
J^Q(X;Y)=\sum_{x,y}P(x)\left[
s_0P(y|x)+\sum_{x_i\in\calX}s_i P(y|x_i)\right]\cdot Q\left(
\frac{t_0P(y|x)+\sum_{x_i\in\calX}t_i P(y|x_i)}
{s_0P(y|x)+\sum_{x_i\in\calX}s_i P(y|x_i)}\right).
\end{equation}
Moreoever, to eliminate the dependence on the specific encoder,
we can think of $\{x_i\}$ as independent 
random variables, take the expectation w.r.t.\ their randomness
(in the same spirit as in \cite{ZZ75}), and obtain
the following information measure
\begin{equation}
\bE\left\{\sum_{x,y}P(x)\left[
s_0P(y|x)+\sum_is_i P(y|X_i)\right]\cdot Q\left(
\frac{t_0P(y|x)+\sum_it_i P(y|X_i)}
{s_0P(y|x)+\sum_{i}s_i P(y|X_i)}\right)\right\},
\end{equation}
where the expectation is w.r.t.\ the product measure of $\{X_i\}$,
$P(x_1,x_2,\ldots)=\prod_iP(x_i)$.
These are the most general information measures,
that obey a data processing inequality, that we can get with
a univariate convex function $Q$. For example, 
returning to eq.\ (\ref{withoutexp}) and taking $s_0=1$, $t_0=0$,
$s_i=sP(x_i)$ ($s \ge 0$, a parameter), and $t_i=P(x_i)$, $x_i\in\calX$,
we have $\mu_0(x,y)=P(x,y)+sP(x)P(y)$,
and $\mu_1(x,y)=P(x)P(y)$, 
and the resulting generalized mutual information
reads
\begin{equation}
\label{simplextension}
J^Q(X;Y)=\sum_{x,y}P(x)[P(y|x)+sP(y)]\cdot
Q\left(\frac{P(y)}{P(y|x)+sP(y)}\right).
\end{equation}
The interesting point concerning these generalized mutual information measures
is that even if we remain in the framework of the 1973 version of the
Ziv--Zakai data processing theorem (as opposed to the 1975 version), we have
added an extra degrees of freedom (in the above example, the parameter $s$),
which may be used in order to improve the obtained bounds. If the inequality
$R^Q(d)\le C^Q$ can be transformed into an inequality on the distortion $d$,
where the lower bound depends on $s$, then this bound can be maximized w.r.t.\
the parameter $s$. If the optimum $s > 0$ yields a distortion bound which is
larger than that of $s=0$, then we have improved on \cite{ZZ73} for the given
choice of the convex function $Q$. Sometimes this optimization may not be
a trivial task, but even if we can just identify one positive value of $s$
(including the limit $s\to\infty$) that is better than $s=0$, then we have
improved on the generalized data processing bound of \cite{ZZ73}, which
corresponds to $s=0$.
This additional degree of freedom may be important,
because, as mentioned in the Introduction, the
variety of convex functions $\{Q\}$ which are convenient to work with, is
somewhat limited (most notably, the functions $Q(z)=z^2$, $Q(z)=1/z$,
$Q(z)=-\sqrt{z}$ and some piecewise linear functions \cite{ZZ73},\cite{ZZ75}).
The next example demonstrates this point.\\

\noindent
{\it Example.}
Consider the information functional (\ref{simplextension}) with the convex
function $Q(z)=-\sqrt{z}$. 
Then, the corresponding generalized mutual information is
\begin{align}
J^Q(U;V)&=-\sum_{u,v}P(u)[P(v|u)+sP(v)]\cdot\sqrt{\frac{P(v)}{P(v|u)+sP(v)}}\nonumber\\
&=-\sum_{u,v}P(u)\sqrt{P(v)[P(v|u)+sP(v)]}\nonumber\\
&=-\sum_{u,v}P(u)P(v)\sqrt{s+\frac{P(v|u)}{P(v)}}.
\end{align}
Consider now the above--described problem of joint source--channel coding, for the following
source and channel:
The source is designated by a
random variable $U$, which is uniformly distributed over the alphabet
$\calU=\{0,1,\ldots,K-1\}$.
The reproduction variable, $V$, takes on values in the
same alphabet, i.e., $\calV=\calU=\{0,1,\ldots,K-1\}$ and the
distortion function is
\begin{equation}
d(u,v)=\left\{\begin{array}{ll}
0 & v=u\\
1 & v=(u+1)\mbox{mod} K\\
\infty & \mbox{elsewhere} \end{array}\right.
\end{equation}
which means that errors other than $v=(u+1)\mbox{mod} K$ are strictly
forbidden. Therefore the channel from $U$ to $V$ must be of the form
\begin{equation}
P(v|u)=\left\{\begin{array}{ll}
1-\epsilon_u & v=u\\
\epsilon_u & v=(u+1)\mbox{mod} K\\
0 & \mbox{elsewhere} \end{array}\right.
\end{equation}
where $\{\epsilon_u\}$ are
parameters taking values in $[0,1]$ and complying with the
distortion constraint
\begin{equation}
\label{distortionconstraint}
\bE\{d(U,V)\}=\frac{1}{K}\sum_{u=0}^{K-1}\epsilon_u\le d.
\end{equation}
The channel is a noise--free $L$--ary channel, i.e.,
its input and output alphabets are $\calX=\calY=\{0,1,\ldots,L-1\}$ with
$P(y|x)=1$ for $y=x$, and $P(y|x)=0$ otherwise.

Obviously, the case $K\le L$ is not interesting because the data can
be conveyed error--free by trivially connecting the source to the
channel. In the other extreme, where $K > 2L$, there must be
some channel input symbol to which at least three source symbols are
mapped. In such a case, it is impossible to avoid at least one of the
forbidden errors in
the reconstruction. Thus, the interesting cases are those for which
$L < K \le 2L$, or equivalently, $\theta\in(1,2]$, where $\theta\dfn K/L$.

We next derive a distortion bound based on the generalized data processing
theorem, in the spirit of \cite{ZZ73} and \cite{ZZ75}, where we now have
the parameter $s$ as a degree of freedom.

As for the source, let us suppose that in addition
to the distortion constraint, we impose
the constraint that the distribution of the reproduction variable $V$, just
like $U$,
must be uniform over its alphabet, namely, $P(v)=1/K$ for all $v\in\calV$.
In this case,
\begin{align}
-J^Q(U;V)&=
\sum_{u,v}P(u)P(v)\sqrt{s+\frac{P(v|u)}{P(v)}}\nonumber\\
&=\frac{1}{K^2}\sum_{u=0}^{K-1}\left[\sqrt{s+K\epsilon_u}+
\sqrt{s+K(1-\epsilon_u)}+(K-2)\sqrt{s}\right]\nonumber\\
&=\frac{1}{K^2}\sum_{u=0}^{K-1}\left[\sqrt{s+K\epsilon_u}+
\sqrt{s+K(1-\epsilon_u)}\right]+\left(1-\frac{2}{K}\right)\sqrt{s}\nonumber\\
&\le\frac{1}{K^2}\cdot K\left[\sqrt{s+Kd}+
\sqrt{s+K(1-d)}\right]+\left(1-\frac{2}{K}\right)\sqrt{s}\nonumber\\
&=\frac{1}{K}\left[\sqrt{s+Kd}+
\sqrt{s+K(1-d)}\right]+\left(1-\frac{2}{K}\right)\sqrt{s},
\end{align}
where the inequality follows from the fact that the maximum of the concave
function
$$\sum_u[\sqrt{s+K\epsilon_u}+\sqrt{s+K(1-\epsilon_u)}],$$
subject to the
distortion constraint (\ref{distortionconstraint}), is achieved when
$\epsilon_u=d$ for all $u\in\calU$.
Thus,
\begin{equation}
R^Q(d)=-\frac{1}{K}\left[\sqrt{s+Kd}-
\sqrt{s+K(1-d)}\right]-\left(1-\frac{2}{K}\right)\sqrt{s}.
\end{equation}

As for the channel, we have:
\begin{align}
-J^Q(X;Y)&=\sum_{x,y}P(x)P(y)\sqrt{s+\frac{P(y|x)}{P(y)}}\nonumber\\
&=\sum_{x'\ne
x}P(x)P(x')\sqrt{s}+\sum_xP^2(x)\sqrt{s+\frac{1}{P(x)}}\nonumber\\
&=\sqrt{s}\left[1-\sum_{x}P^2(x)\right]+\sum_xP^2(x)\sqrt{s+\frac{1}{P(x)}}\nonumber\\
&=\sqrt{s}+\sum_xP^2(x)\left(\sqrt{s+\frac{1}{P(x)}}-\sqrt{s}\right)\nonumber\\
&=\sqrt{s}+\sum_xP^2(x)\cdot\frac{1/P(x)}{\sqrt{s+1/P(x)}+\sqrt{s}}\nonumber\\
&=\sqrt{s}+\sum_x\frac{P(x)}{\sqrt{s+1/P(x)}+\sqrt{s}}.
\end{align}
The function $f(t)=t/[\sqrt{s+1/t}+\sqrt{s}]$ 
is convex in $t$ (for fixed $s$)
since $f''(t)\ge 0$ for all
$t\ge 0$, as can readily be verified.
Thus, $-J^Q(X;Y)$ is
minimized by the uniform distribution $P(x)=1/L$, $\forall x$, which leads to
the `capacity' expression:
\begin{equation}
C^Q=-\sqrt{s}-\frac{1}{\sqrt{s}+\sqrt{s+L}}.
\end{equation}
Applying now the data processing theorem,
\begin{equation}
R^Q(d)\le C^Q,
\end{equation}
we obtain, after rearranging terms
\begin{equation}
\sqrt{s+Kd}+
\sqrt{s+K(1-d)}\ge \frac{K}{\sqrt{s}+\sqrt{s+L}}+2\sqrt{s}.
\end{equation}
Squaring both sides, we have:
\begin{equation}
2s+K+2\sqrt{(s+Kd)[s+K(1-d)]}\ge
\left[\frac{K}{\sqrt{s}+\sqrt{s+L}}+2\sqrt{s}\right]^2
\end{equation}
or
\begin{equation}
2\sqrt{(s+Kd)[s+K(1-d)]}\ge
\left[\frac{K}{\sqrt{s}+\sqrt{s+L}}+2\sqrt{s}\right]^2
-2s-K,
\end{equation}
which after squaring again and applying some further straightforward
algebraic manipulations, gives eventually the following inequality on
the distortion $d$:
\begin{equation}
4d(1-d)\ge \psi(s),
\end{equation}
where
\begin{equation}
\psi(s)\dfn\frac{1}{K^2}
\left[\left(\frac{K}{\sqrt{s}+\sqrt{s+L}}+2\sqrt{s}\right)^2-2s-K\right]^2-\frac{4s(s+K)}{K^2}.
\end{equation}
The resulting lower bound on the distortion is the smaller of the
two solutions of the equation $4d(1-d)=\psi(s)$, which is
\begin{equation}
d_s\dfn \frac{1}{2}-\frac{1}{2}\sqrt{1-\psi(s)}.
\end{equation}
Thus, the larger is $\psi(s)$, the better is the bound.
The choice $s=0$, which corresponds to the usual Ziv--Zakai bound for
$Q(z)=-\sqrt{z}$, yields
\begin{equation}
\psi(0)=
\frac{1}{K^2}\left[\left(\frac{K}{\sqrt{L}}\right)^2-K\right]^2=\left(\frac{K}{L}-1\right)^2=(\theta-1)^2.
\end{equation}
However, it turns out that $s=0$ is not the best choice of $s$.
We next examine the limit $s\to\infty$. To this end, we derive a lower bound
to $\psi(s)$ which is more convenient to analyze in this limit.
Note that for $s\ge L/8$, it is guaranteed that the expression in the square
brackets of the expression defining $\psi(s)$, is positive, which means
that an upper bound on $\sqrt{s+L}$ would yield a lower bound to $\psi(s)$.
Thus,  upper bounding $\sqrt{s+L}$ by
$$\sqrt{s+L}=\sqrt{s}\cdot\sqrt{1+L/s}\le\sqrt{s}\left(1+\frac{L}{2s}\right),$$
we get
\begin{align}
K^2\psi(s)&=
\left[\left(\frac{K}{\sqrt{s}+\sqrt{s+L}}+2\sqrt{s}\right)^2-2s-K\right]^2-4s^2-4Ks\nonumber\\
&\ge\left[\left(\frac{K}{\sqrt{s}(2+L/2s)}+2\sqrt{s}\right)^2-2s-K\right]^2-4s^2-4Ks\nonumber\\
&=K^2\left(\frac{4s-L}{4s+L}\right)^2+\frac{16K^4s^2}{(4s+L)^4}-\frac{8KLs}{4s+L}+
\frac{16K^2s^2}{(4s+L)^2}+\frac{8K^3s(4s-L)}{(4s+L)^3}\nonumber\\
&\dfn K^2\psi_0(s),
\end{align}
where between the second and the third lines, we have skipped some standard
algebraic operations. Taking now the limit $s\to\infty$, we obtain
\begin{equation}
\psi_\infty=\lim_{s\to\infty}\psi_0(s)=\frac{1}{K^2}(K^2+0-2KL+K^2+0)=2\left(1-\frac{L}{K}\right)=
2\left(1-\frac{1}{\theta}\right),
\end{equation}
which yields a better bound than the bound of $s=0$ since
\begin{equation}
2\left(1-\frac{1}{\theta}\right)> (\theta-1)^2 
\end{equation}
for all $\theta\in(1,2)$.

It is interesting to compare this also to the classical data processing
theorem: Since
\begin{equation}
R(d)=\log K-h_2(d) 
\end{equation}
and
\begin{equation}
C=\log L, 
\end{equation}
then the ordinary data
processing theorem yields the bound
\begin{equation}
h_2(d)\ge\log\theta.
\end{equation}
Since 
\begin{equation}
h_2(d)\ge 4d(1-d) 
\end{equation}
and 
\begin{equation}
2\left(1-\frac{1}{\theta}\right)\ge \log_2\theta 
\end{equation}
within the relevant range of $\theta$, the bound
pertaining to $s\to\infty$ is also better than the classical bound for this
case. This completes the description of the example. $\Box$

Finally, we should comment that the monotonicity result concerning $V(t)$
contains as special cases, not only
the H--theorem,
as well as all other earlier mentioned monotonicity results,
but also the 1975 Zakai--Ziv generalized data processing \cite{ZZ75}.
Consider a Markov chain $U\to V\to W$, where $U$, $V$
and $W$ are random variables that take on values in (finite) alphabets,
$\calU$, $\calV$, and $\calW$, respectively.
Let us now map between the Markov chain $(U,V,W)$ and the Markov process
$\{X_t\}$
in the following manner:
$(u,v)\in\calU\in\calV$ is assigned to the state $x'$ of the process at time
$t$, whereas
$(u,w)\in\calU\in\calW$ corresponds\footnote{While $\calV$ and $\calW$ may be
different (finite) alphabets, $x$ and $x'$, of the original Markov process,
must taken on values in the same
alphabet. Assuming, without loss of generality, that
$\calV=\{1,2,\ldots,|\calV|\}$ and
$\calW=\{1,2,\ldots,|\calW|\}$, then for the purpose of
this mapping, we can unify these alphabets to be both
$\{1,2,\ldots,\max\{|\calV|,|\calW|\}\}$ and complete the missing
elements of the extended transition
matrix $P(w|v)$ in a consistent manner,
according to the actual support of each
distribution. We omit
further technical details herein.}
to $x$ at time $t+1$. Now, defining accordingly,
\begin{eqnarray}
\mu_t^{0}(x')&=&P(u,v),\\
\mu_t^{1}(x')&=&P(u)P(v),\\
\mu_{t+1}^{0}(x)&=&P(u,w),
\end{eqnarray}
and
\begin{equation}
\mu_{t+1}^{1}(x)=P(u)P(w),
\end{equation}
then due to the Markov property
of $(U,V,W)$, both measures satisfy the recursion with
$P(w|v)$ playing the role\footnote{Consider the component $u$ of $x'=(u,v)$
and
$x=(u,w)$ simply as an index.}
of $P(x|x')$. I.e.,
\begin{align}
P(u,w)&\dfn \mu_{t+1}^{0}(x)\nonumber\\
&=\sum_{x'}\mu_t^{0}(x')P(x|x')\nonumber\\
&=\sum_vP(u,v)P(w|v)
\end{align}
and
\begin{align}
P(u)P(w)&\dfn \mu_{t+1}^{1}(x)\nonumber\\
&=\sum_{x'}\mu_t^{1}(x')P(x|x')\nonumber\\
&=\sum_vP(u)P(v)P(w|v)
\end{align}
Thus, for $Q(z)=-\ln z$, the monotonicity of $V(t)$ is nothing but the
data processing of the classical mutual information. For a general
function $Q$ of one variable ($k=1$), this gives the generalized
data processing theorem of \cite{ZZ73}. Furthermore, letting $Q$ be a general convex function
of $k$ variables,
and $\mu_t^{0}(x')=P(u,v)$ as before, we get the more general
form of the data processing inequality of \cite{ZZ75}.

The above extension of the H--theorem gives rise to a seemingly more
general data processing theorem than in \cite{ZZ75},
as it is not necessary to let $\mu_t^{0}(x)$ be the actual joint probability
distribution. However, when looking at the entire class of convex functions
with an arbitrary number of arguments,
this is not really more general, as the corresponding
generalized mutual information can readily be transformed back to the
form of the 1975 Zakai--Ziv information functional using again the perspective
operation. Indeed, as mentioned in the Introduction and 
shown in \cite[Theorem 7.1]{ZZ75}, the class of
generalized mutual information measures studied therein cannot be improved
upon in the sense that there always exist choices of $Q$ and $\{\mu_i\}$ that
provide tight bounds on the distortion of the optimum system.

\section{Summary and Conclusion}
\label{summary}

The main contributions of this work can be summarized as follows:
First, we have establisehd a unified framework and a relationship between (a generalized version of) 
the second law of thermodynamics and the generalized data processing theorems
of Zakai and Ziv. This unified framework turns out to strengthen and expand
both of these pieces of theory: Concerning the second law of thermodynamics,
we have identified a significantly more general information measure, which is
a monotonic function of time, when it operates on a Markov process. As for
the generalized Ziv--Zakai data processing theorem, we have proposed a wider
class of information measures obeying the data processing theorem, which
includes free parameters that may be optimized so as to tighten the distortion
bounds.

\section*{Acknowledgment}

Interesting discussions with J.~Ziv and M.~Zakai are acknowledged with thanks.

\end{document}

%% file: mm1.pstex_t
\begin{picture}(0,0)%
\includegraphics{mm1.pstex}%
\end{picture}%
\setlength{\unitlength}{3947sp}%
\begingroup\makeatletter\ifx\SetFigFont\undefined%
\gdef\SetFigFont#1#2#3#4#5{%
  \reset@font\fontsize{#1}{#2pt}%
  \fontfamily{#3}\fontseries{#4}\fontshape{#5}%
  \selectfont}%
\fi\endgroup%
\begin{picture}(5373,1499)(341,-1071)
\put(823,334){\makebox(0,0)[lb]{\smash{{\SetFigFont{14}{16.8}{\rmdefault}{\mddefault}{\itdefault}{$\lambda$}%
}}}}
\put(2192,334){\makebox(0,0)[lb]{\smash{{\SetFigFont{14}{16.8}{\rmdefault}{\mddefault}{\itdefault}{$\lambda$}%
}}}}
\put(3496,334){\makebox(0,0)[lb]{\smash{{\SetFigFont{14}{16.8}{\rmdefault}{\mddefault}{\itdefault}{$\lambda$}%
}}}}
\put(4735,334){\makebox(0,0)[lb]{\smash{{\SetFigFont{14}{16.8}{\rmdefault}{\mddefault}{\itdefault}{$\lambda$}%
}}}}
\put(1018,-1036){\makebox(0,0)[lb]{\smash{{\SetFigFont{14}{16.8}{\rmdefault}{\mddefault}{\itdefault}{$\mu$}%
}}}}
\put(2257,-1036){\makebox(0,0)[lb]{\smash{{\SetFigFont{14}{16.8}{\rmdefault}{\mddefault}{\itdefault}{$\mu$}%
}}}}
\put(3562,-1036){\makebox(0,0)[lb]{\smash{{\SetFigFont{14}{16.8}{\rmdefault}{\mddefault}{\itdefault}{$\mu$}%
}}}}
\put(4866,-1036){\makebox(0,0)[lb]{\smash{{\SetFigFont{14}{16.8}{\rmdefault}{\mddefault}{\itdefault}{$\mu$}%
}}}}
\put(562,-319){\makebox(0,0)[lb]{\smash{{\SetFigFont{14}{16.8}{\rmdefault}{\mddefault}{\itdefault}{0}%
}}}}
\put(1801,-319){\makebox(0,0)[lb]{\smash{{\SetFigFont{14}{16.8}{\rmdefault}{\mddefault}{\itdefault}{1}%
}}}}
\put(3105,-384){\makebox(0,0)[lb]{\smash{{\SetFigFont{14}{16.8}{\rmdefault}{\mddefault}{\itdefault}{2}%
}}}}
\put(4409,-384){\makebox(0,0)[lb]{\smash{{\SetFigFont{14}{16.8}{\rmdefault}{\mddefault}{\itdefault}{3}%
}}}}
\put(5714,-384){\makebox(0,0)[lb]{\smash{{\SetFigFont{14}{16.8}{\rmdefault}{\mddefault}{\itdefault}{$\cdot\cdot\cdot$}%
}}}}
\end{picture}%

%% file: circuit.pstex_t
\begin{picture}(0,0)%
\includegraphics{circuit.pstex}%
\end{picture}%
\setlength{\unitlength}{3947sp}%
\begingroup\makeatletter\ifx\SetFigFont\undefined%
\gdef\SetFigFont#1#2#3#4#5{%
  \reset@font\fontsize{#1}{#2pt}%
  \fontfamily{#3}\fontseries{#4}\fontshape{#5}%
  \selectfont}%
\fi\endgroup%
\begin{picture}(5537,1449)(173,-748)
\put(2821,382){\makebox(0,0)[lb]{\smash{{\SetFigFont{6}{7.2}{\rmdefault}{\mddefault}{\itdefault}{1}%
}}}}
\put(3588,273){\makebox(0,0)[lb]{\smash{{\SetFigFont{6}{7.2}{\rmdefault}{\mddefault}{\itdefault}{2}%
}}}}
\put(4289,382){\makebox(0,0)[lb]{\smash{{\SetFigFont{6}{7.2}{\rmdefault}{\mddefault}{\itdefault}{3}%
}}}}
\put(5056,338){\makebox(0,0)[lb]{\smash{{\SetFigFont{6}{7.2}{\rmdefault}{\mddefault}{\itdefault}{4}%
}}}}
\put(5056,-407){\makebox(0,0)[lb]{\smash{{\SetFigFont{6}{7.2}{\rmdefault}{\mddefault}{\itdefault}{5}%
}}}}
\put(4289,-341){\makebox(0,0)[lb]{\smash{{\SetFigFont{6}{7.2}{\rmdefault}{\mddefault}{\itdefault}{6}%
}}}}
\put(3544,-451){\makebox(0,0)[lb]{\smash{{\SetFigFont{6}{7.2}{\rmdefault}{\mddefault}{\itdefault}{7}%
}}}}
\put(3106,470){\makebox(0,0)[lb]{\smash{{\SetFigFont{6}{7.2}{\rmdefault}{\mddefault}{\itdefault}{$R_{12}$}%
}}}}
\put(3829,470){\makebox(0,0)[lb]{\smash{{\SetFigFont{6}{7.2}{\rmdefault}{\mddefault}{\itdefault}{$R_{23}$}%
}}}}
\put(4553,470){\makebox(0,0)[lb]{\smash{{\SetFigFont{6}{7.2}{\rmdefault}{\mddefault}{\itdefault}{$R_{34}$}%
}}}}
\put(5100,-35){\makebox(0,0)[lb]{\smash{{\SetFigFont{6}{7.2}{\rmdefault}{\mddefault}{\itdefault}{$R_{45}$}%
}}}}
\put(4574,-276){\makebox(0,0)[lb]{\smash{{\SetFigFont{6}{7.2}{\rmdefault}{\mddefault}{\itdefault}{$R_{56}$}%
}}}}
\put(3808,-276){\makebox(0,0)[lb]{\smash{{\SetFigFont{6}{7.2}{\rmdefault}{\mddefault}{\itdefault}{$R_{67}$}%
}}}}
\put(3632,  9){\makebox(0,0)[lb]{\smash{{\SetFigFont{6}{7.2}{\rmdefault}{\mddefault}{\itdefault}{$R_{27}$}%
}}}}
\put(2953,119){\makebox(0,0)[lb]{\smash{{\SetFigFont{6}{7.2}{\rmdefault}{\mddefault}{\itdefault}{$P_1$}%
}}}}
\put(3610,601){\makebox(0,0)[lb]{\smash{{\SetFigFont{6}{7.2}{\rmdefault}{\mddefault}{\itdefault}{$P_2$}%
}}}}
\put(4421,119){\makebox(0,0)[lb]{\smash{{\SetFigFont{6}{7.2}{\rmdefault}{\mddefault}{\itdefault}{$P_3$}%
}}}}
\put(5144,492){\makebox(0,0)[lb]{\smash{{\SetFigFont{6}{7.2}{\rmdefault}{\mddefault}{\itdefault}{$P_4$}%
}}}}
\put(5144,-561){\makebox(0,0)[lb]{\smash{{\SetFigFont{6}{7.2}{\rmdefault}{\mddefault}{\itdefault}{$P_5$}%
}}}}
\put(4421,-561){\makebox(0,0)[lb]{\smash{{\SetFigFont{6}{7.2}{\rmdefault}{\mddefault}{\itdefault}{$P_6$}%
}}}}
\put(3347,-539){\makebox(0,0)[lb]{\smash{{\SetFigFont{6}{7.2}{\rmdefault}{\mddefault}{\itdefault}{$P_7$}%
}}}}
\put(250,391){\makebox(0,0)[lb]{\smash{{\SetFigFont{6}{7.2}{\rmdefault}{\mddefault}{\itdefault}{1}%
}}}}
\put(780,391){\makebox(0,0)[lb]{\smash{{\SetFigFont{6}{7.2}{\rmdefault}{\mddefault}{\itdefault}{2}%
}}}}
\put(1344,391){\makebox(0,0)[lb]{\smash{{\SetFigFont{6}{7.2}{\rmdefault}{\mddefault}{\itdefault}{3}%
}}}}
\put(1940,358){\makebox(0,0)[lb]{\smash{{\SetFigFont{6}{7.2}{\rmdefault}{\mddefault}{\itdefault}{4}%
}}}}
\put(1973,-372){\makebox(0,0)[lb]{\smash{{\SetFigFont{6}{7.2}{\rmdefault}{\mddefault}{\itdefault}{5}%
}}}}
\put(1245,-372){\makebox(0,0)[lb]{\smash{{\SetFigFont{6}{7.2}{\rmdefault}{\mddefault}{\itdefault}{6}%
}}}}
\put(714,-372){\makebox(0,0)[lb]{\smash{{\SetFigFont{6}{7.2}{\rmdefault}{\mddefault}{\itdefault}{7}%
}}}}
\end{picture}%